\documentstyle[12pt]{article}
\title{\bf Exact solutions of Dirac equation on
(1+1)-dimensional spacetime coupled to a static scalar field}
\author{F. Darabi$^{a}$\thanks{e-mail:
f.darabi@azaruniv.edu}, S. K. Moayedi$^{b}$\thanks{e-mail:
S-MOAYEDI@araku.ac.ir}, A. R. Ahmadi$^{b}$\\
$^{a}${\small Department of Physics, Azarbaijan University of
Tarbiat Moallem , Tabriz,  Iran.}\\
$^{b}${\small Department of Physics, Arak University, Arak, Iran.}
}
\begin{document}
\maketitle \baselineskip.4in
\begin{abstract}
We use a generalized scheme of supersymmetric quantum mechanics to
obtain the energy spectrum and wave function for Dirac equation in
(1+1)-dimensional spacetime coupled to a static scalar field.
\end{abstract}
\newpage

\section{Introduction}

In the mathematical framework of two dimensional physics the exact
solutions of Klein-Gordon and Dirac equations and their spectrum
on a (1+1)-dimensional background is of particular interest. A lot
of studies have been done on the motion of Klein-Gordon particle
\cite{1, 2}, as well as fermionic one \cite{3, 4} on the (1+1)-
dimensional manifolds.

In our previous works \cite{5}, we have exactly solved the
Klein-Gordon equation on a static (1+1)-dimensional space-time,
and Dirac equation on a (1+1)-dimensional gravitational
background, by using of the {\it standard} techniques of
supersymmetric quantum mechanics. In this paper, we use a {\it
generalized} scheme of supersymmetric quantum mechanics (SUSY QM)
to obtain the exact solution and energy spectrum of Dirac equation
coupled to a static scalar field in (1+1) dimensions. It is shown
that in a special limiting case, the Dirac spinor obtained by the
generalized SUSY is reduced to the spinor obtained by the standard
SUSY.

\section{Massless Dirac equation coupled to a static scalar field}

On a (1+1)-dimensional Lorentzian manifold with the signature (+1,
-1), the massless Dirac equation coupled to a static scalar field
$\phi(x)$ is written as \cite{7}
\begin{equation}
[i\gamma^{\mu}\partial_\mu-\phi(x)]\Psi(x, t)=0,
\label{1}
\end{equation}
where the matrices $\gamma^\mu$ are the generators of Clifford
algebra of two-dimensional flat space-time
\begin{equation}
\{\gamma^\mu, \gamma^\nu\}=2\eta^{\mu \nu}.
\label{2}
\end{equation}
Following Jackiw and Rebbi \cite{6}, we take the following
representations for the $\gamma^\mu$ matrices
\begin{equation}
\gamma^0=\sigma^1\:\:\:,\:\:\:\gamma^1=i\sigma^3,
\label{3}
\end{equation}
where $\sigma^1, \sigma^3$ are the Pauli spin matrices. We
consider $\Psi$ as a two component spinor
\begin{equation}
\Psi(x, t)= e^{-i\epsilon t}\left(\begin{array}{c} \Psi^{(1)}(x) \\
\Psi^{(2)}(x)
\end{array}\right)_.
\label{4}
\end{equation}
Now, the Dirac equation (\ref{1}) becomes
\begin{equation}
\left \{ \begin{array}{ll}
(\frac{d}{dx}+\phi(x))\Psi_n^{(1)}(x)=\epsilon_n \Psi_n^{(2)}(x),\\
\\
(-\frac{d}{dx}+\phi(x))\Psi_n^{(2)}(x)=\epsilon_n \Psi_n^{(1)}(x).
\end{array}\right.
\label{5}
\end{equation}
If we define the operators
\begin{equation}
\left\{ \begin{array}{ll}
A=\frac{d}{dx}+\phi(x),\\
\\
A^\dag=-\frac{d}{dx}+\phi(x),
\end{array} \right.
\label{6}
\end{equation}
then, the equations (\ref{5}) are written as
\begin{equation}
\left\{\begin{array}{ll}
A\Psi_n^{(1)}(x)=\epsilon_n \Psi_n^{(2)}(x),\\
\\
A^\dag \Psi_n^{(2)}(x)=\epsilon_n \Psi_n^{(1)}(x).
\end{array}\right.
\label{7}
\end{equation}
By operating $A^\dag$ and $A$, from left, on both sides of the
first and second equations in Eq.(\ref{7}), respectively, we
obtain the following supersymmetric equations
\begin{equation}
\left\{ \begin{array}{ll}
H_1\Psi_n^{(1)}(x)=\epsilon^2_n\Psi_n^{(1)}(x)=E_n^{(1)}\Psi_n^{(1)}(x),\\
\\
H_2\Psi_n^{(2)}(x)=\epsilon^2_n\Psi_n^{(2)}(x)=E_n^{(1)}\Psi_n^{(2)}(x),
\end{array}\right.
\label{8}
\end{equation}
where $E_n^{(1)}=E_n^{(2)}=\epsilon_n^2>0$ which is commonly
called the {\it broken supersymmetry} \cite{7}. To solve
Eqs.(\ref{8}) one may use the standard technics of supersymmetric
quantum mechanics.

\section{Exact solutions of 2D Dirac equation by generalized SUSY }

We define the generalized operators \cite{8}
\begin{equation}
\left \{ \begin{array}{ll}
B=\frac{d}{dx}+\Phi(x),\\
\\
B^\dag=-\frac{d}{dx}+\Phi(x),
\end{array}\right.
\label{9}
\end{equation}
where $\Phi(x)$ is called the generalized {\it superpotential} and
has the form
\begin{equation}
\Phi(x)=\phi(x)+g(x).
\label{10}
\end{equation}
The approach of generalized supersymmetric quantum mechanics is to
suppose that the partner Hamiltonian $H_2$ is not uniquely
decomposed as $AA^\dag$. One may then look for a suitable
superpotential $\Phi(x)$ so that
\begin{equation}
\tilde{H}_2=BB^\dag=AA^\dag=H_2.
\label{11}
\end{equation}
The above equation means that both $H_2$ and $\tilde{H}_2$ have
the same energy spectrum and wave function. Now, by inserting
Eqs.(\ref{6}) and (\ref{9}) in Eq.(\ref{11}) we find that $g(x)$
must satisfy the following Riccati type equation
\begin{equation}
g^\prime(x)+2\phi(x)g(x)+g^2(x)=0.
\label{12}
\end{equation}
The partner Hamiltonian of $\tilde{H}_2$ is now defined as
$\tilde{H}_1=B^\dag B$ which is related to $H_1$ as follows
\begin{equation}
\tilde{H}_1=H_{1}-2g^\prime(x).
\label{13}
\end{equation}
Dirac equation in the presence of the generalized background
scalar field $\Phi(x)$ is as follows
\begin{equation}
[i\gamma^{\mu}\partial_\mu-\Phi(x)]\tilde{\Psi}(x, t)=0.
\label{14}
\end{equation}
Now, by assuming the generalized spinor
\begin{equation}
\tilde{\Psi}(x, t)= e^{-i\epsilon t}\left(\begin{array}{c} \tilde{\Psi}^{(1)}(x) \\
\tilde{\Psi}^{(2)}(x)
\end{array}\right),
\label{15}
\end{equation}
and using (\ref{3}) we obtain
\begin{equation}
\left \{ \begin{array}{ll}
\tilde{H}_1\tilde{\Psi}_n^{(1)}(x)=\epsilon^2_n\tilde{\Psi}_n^{(1)}(x),\\
\\
\tilde{H}_2\tilde{\Psi}_n^{(2)}(x)=\epsilon^2_n\tilde{\Psi}_n^{(2)}(x),
\end{array}\right.
\label{16}
\end{equation}
which are to be solved to construct the Dirac spinor (\ref{15}).
Now, we use the above formalism to solve exactly the Dirac
equation for a typical example.

Consider the following superpotential
\begin{equation}
\phi(x)=2\tanh(x).
\label{17}
\end{equation}
By using of the Riccati equation (\ref{12}), the function $g(x)$
for this superpotential is obtained
\begin{equation}
g(x)=\frac{3\cosh^{-4}(x)}{4\lambda+3\tanh(x)-\tanh^3(x)+2},
\label{18}
\end{equation}
where the real constant $\lambda$ must be so chosen that the
denominator does not vanish. If we now use the change of variable
$y=\sinh(x)$, then the wave functions and the energy spectrum are
obtained by the generalized SUSY as
\begin{equation}
\tilde{\Psi}_n^{(1)}(y)=\frac{2in}{1+y^2}P_n^{(-\frac{5}{2},
-\frac{5}{2})}(iy)+\frac{3P_{n-1}^{(-\frac{3}{2},
-\frac{3}{2})}(iy)}{(1+y^2)[y(3+2y^2)+(4\lambda+2)(1+y^2)^{\frac{3}{2}}]},
\label{19}
\end{equation}
\begin{equation}
\tilde{\Psi}_n^{(2)}(y)=(1+y^2)^{-\frac{1}{2}}P_{n-1}^{(-\frac{3}{2},
-\frac{3}{2})}(iy), \label{20}
\end{equation}
\begin{equation}
\epsilon^2_n=n(4-n).
\label{21}
\end{equation}
Whereas, using the standard SUSY we obtain
\begin{equation}
\Psi_n^{(1)}(y)=(1+y^2)^{-1}P_{n}^{(-\frac{5}{2},
-\frac{5}{2})}(iy),
\label{22}
\end{equation}
\begin{equation}
\Psi_n^{(2)}(y)=\tilde{\Psi}_n^{(2)}(y),
\label{23}
\end{equation}
\begin{equation}
\epsilon^2_n=n(4-n).
\label{24}
\end{equation}
It is to be noted that $P_n^{(\alpha, \beta)}(x)$ are the Jacobi
polynomials with the integer $n$. Now we discuss about the allowed
modes of the energy spectrum in Eq.(\ref{21}).

Since one of the major features of the broken generalized
supersymmetry is the positivity of the energy spectrum for the
partner Hamiltonians $\tilde{H}_1$ and $\tilde{H}_2$, then the
allowed values for the integer $n$ are $n=1, 2, 3, $... which lead
to the allowed energy levels for Dirac particle. One may also show
that considering the limiting condition $\lambda \rightarrow
\infty$, the Dirac spinor obtained by Eqs.(\ref{19}), (\ref{20})
is reduced to the one obtained by Eqs.(\ref{22}), (\ref{23}).

\section*{Conclusion}

In this letter we have shown that by obtaining the energy spectrum
and the wave function corresponding to the Dirac equation
(\ref{1}) in the presence of superpotential (\ref{17}), and using
the non-uniqueness decomposition of $H_2$, namely Eq.(\ref{11}),
together with Eqs.(\ref{10}), (\ref{12}) a generalized
superpotential is obtained for which the wave functions and the
energy spectrum for the Dirac equation (\ref{14}) are obtained as
Eqs.(\ref{19}), (\ref{20}) and (\ref{21}). It seems this procedure
works as well for other potentials for which the Dirac equation
(\ref{1}) has analytic solution.

\end{document}